\documentclass[useAMS,usenatbib]{mnras}

\usepackage{amssymb}
\usepackage[tbtags]{amsmath}
\usepackage{amsfonts}
\usepackage{graphicx}
\usepackage{subfig}
\usepackage{color}
\usepackage[table]{xcolor}
\usepackage{multirow}
\usepackage{tikz}
\usepackage{color, colortbl}
\usepackage{times}
\definecolor{Gray}{gray}{0.5}
\usepackage{tabularx}
\hypersetup{draft}

\def\aj{AJ}
\def\araa{ARA\&A}
\def\apj{ApJ}
\def\apjl{ApJ}
\def\apjs{ApJS}
\def\aap{A\&A}
\def\mnras{MNRAS}
\def\nat{Nature}
\def\pasp{PASP} 

\def\FeH{\mathrm{[Fe/H]}}
\def\aFe{[\alpha/\mathrm{Fe}]}
\def\MH{[\mathrm{M}/\mathrm{H}]}
\def\aM{[\alpha/\mathrm{M}]}

\newcommand{\diff}{\mathrm{d}}

{\newif\ifnotend
\notendtrue
\def\veclist{ABCDEFGHIJKLMNOPQRSTUVWXYZabcdefghijklmnopqrstuvwxyz.}
\def\top#1#2.{#1}
\def\tail#1#2.{#2.}
\loop\expandafter\xdef\csname v\expandafter\top\veclist\endcsname%
{{\noexpand\bf\expandafter\top\veclist}}
\edef\veclist{\expandafter\tail\veclist}
\if\veclist.\notendfalse\fi\ifnotend\repeat}

\title[A spectroscopic Mass, Distance, and Age Estimator]{MADE: A spectroscopic Mass, Age, and Distance Estimator for red giant stars with Bayesian machine learning}

\author[Payel Das and Jason L. Sanders]{Payel Das$^{1}$\thanks{E-mail:payel.das@physics.ox.ac.uk} 
and {Jason L. Sanders$^{2}$}\\
$^{1}$Rudolf Peierls Centre for Theoretical Physics, University of Oxford, OX1 3NP, UK\\
$^{2}$Institute of Astronomy, University of Cambridge, Madingley Road, Cambridge, CB3 0HA, UK}

\begin{document}

\pagerange{\pageref{firstpage}--\pageref{lastpage}} \pubyear{2015}

\maketitle

\label{firstpage}

\begin{abstract} 
We present a new approach (MADE) that generates mass, age, and distance estimates of red giant stars from a combination of astrometric, photometric, and spectroscopic data. The core of the approach is a Bayesian artificial neural network (ANN) that learns from and completely replaces stellar isochrones. The ANN is trained using a sample of red giant stars with mass estimates from asteroseismology. A Bayesian isochrone pipeline uses the astrometric, photometric, spectroscopic, and asteroseismology data to determine posterior distributions for the training outputs: mass, age, and distance. Given new inputs, posterior predictive distributions for the outputs are computed, taking into account both input uncertainties, and uncertainties in the ANN parameters. 

We apply MADE to $\sim10\,000$ red giants in the overlap between the 14$^{\mathrm{th}}$ data release from the APO Galactic Evolution Experiment \citep[APOGEE, ][]{abol+18} and the Tycho-Gaia astrometric solution \citep[TGAS, ][]{mich+15}. The ANN is able to reduce the uncertainty on mass, age, and distance estimates for training-set stars with high output uncertainties allocated through the Bayesian isochrone pipeline. The fractional uncertainties on mass are $<10\%$  and on age are between $10$ to $25\%$. Moreover, the time taken for our ANN to predict masses, ages, and distances for the entire catalogue of APOGEE-TGAS stars is of a similar order of the time taken by the Bayesian isochrone pipeline to run on a handful of stars. Our resulting catalogue clearly demonstrates the expected thick and thin disc components in the $\MH$-$\aM$ plane, when examined by age.
\end{abstract}


\section[]{Introduction} 

Knowledge of the distance, age, and chemical composition of stars is fundamental for understanding the Galaxy's history of formation, enrichment, and dynamical evolution \citep{bland+16}. When interpreted through chemodynamical models such as `extended distribution functions' \citep{sanders+15,das+16a}, they can constrain Galactic evolution as well as the gravitational potential and dark matter content of the Milky Way. However, distance, age, and chemical composition are not measured directly and need to be inferred from stellar models. 

The oldest method of measuring distance is through the measurement of a star's parallax. This method has traditionally been limited to local Galactic studies, but the second data release from the Gaia mission \citep{gaia+16,gaia+18} has provided parallaxes for a billion stars down to 20$^{\mathrm{th}}$ magnitude. Variable stars, such as Cepheids, can be used as standard candles and so their apparent magnitudes can be mapped to a narrow range of intrinsic luminosities, allowing distances to be easily derived. This method can be used to large distances but only with a limited range of stellar types. Another example of a model-independent method estimates distances for spectroscopic twins of nearby stars with known parallaxes \citep{jofre+15,jofre+17}. This method does not degrade with increasing stellar distance, but is limited by the availability of high-resolution stellar spectra and parallaxes for a range of stellar types. 

The age of an individual star can be estimated through a range of almost entirely model-dependent methods \citep[for a review see ][]{soderblom+10}. \cite{pont+04} developed a Bayesian method to estimates the ages of dwarf stars from temperature, luminosity and metallicity data using isochrones. Several other authors \citep[e.g.][]{burnett+10,carlin+15, santiago+16} have since developed similar procedures to estimate the mass, age, distance, and metallicity of a star from photometric and spectroscopic data, and where available, luminosity. Parallaxes, which coupled with photometric data lay strong constraints on the intrinsic luminosity, have also been incorporated into these pipelines where available \citep[e.g.][]{jorg+05,dasilva+06,mcmillan+17,mints+18,queiroz+18}. Without an indication of the star's intrinsic luminosity from variable star relations or parallax, the inferred ages are only reliable in the turn-off region. The approach works well for stellar parameter estimation over a large range of distance.

In recent years, asteroseismology has revealed considerable potential in determining ages for giant stars. Surveys such as the Microvariablity and Oscillations of Stars (MOST) mission \citep{guenther+05}, the Convection, Rotation et Transits planétaires (CoRot) mission \citep{deridder+09}, and Kepler \citep{bedding+10} have been extremely successful in probing the stellar interiors for red giants \citep[e.g., ][]{chaplin+13,silva+15}. These missions take high-cadence, high-precision stellar photometry that show oscillation modes in Fourier space. The frequency of maximum intensity, $\nu_{\mathrm{max}}$, and the frequency spacing of the modes, $\Delta\nu$, are related to the density and mass of the stars through simple scaling relations \citep{ulrich+86,kjeldsen+95}. As giants spend most of their life in the main sequence phase, the age at which they entered the giant branch is approximately their present age. This age is fixed by the mass of the star i.e. their mass is a strong indicator of their age. Determination of ages for red giants with asteroseismology data has already been demonstrated to be a powerful probe of Milky Way chemodynamics, in particular when combined with spectroscopy \citep{miglio+13,rodrigues+14,stello+15,casagrande+16,anders+17}.

At present, these asteroseismological surveys have yielded thousands of masses rather than millions. However, they have made it possible to determine the significance of mass-driven differences in the spectra of giants. This has significantly broadened the scope for asteroseismology through the development of `spectroscopic mass estimators', which use spectral parameters or spectra to predict mass $m$.
\cite{masseron+15} showed that masses of red giants could be derived from spectra given the [C/N] change due to dredge up in the giant phase. \cite{martig+16} developed simple linear regression models to estimate masses of giants from $\log g$, $T_{\mathrm{eff}}$, $\FeH$, carbon abundance, and nitrogen abundance using data from the APOGEE-Kepler
Asteroseismology Science Consortium
(APOKASC) catalogue \citep{pinson+14}. This catalogue contains stars in the cross-match between stars with spectroscopic data from APO Galactic Evolution Experiment \citep[APOGEE, ][]{majewski+17} and asteroseismology from the Kepler mission \citep{haas+10}. \cite{ness+16} trained the {\it Cannon} \citep{ness+15} on this catalogue, a regression model that fits each pixel of a star's spectrum with a polynomial function of the spectroscopic labels ($\log g$, $T_{\mathrm{eff}}$, and $\FeH$, $\aFe$, $m$). The mass estimates can then be combined with isochrones to infer ages for all $\sim60\,000$ stars in APOGEE DR12 \citep{alam+15}, rather than just the $\sim5000$ in APOKASC, clearly revealing relations such as variation of age with Galactic position in mono-abundance populations.   

The potential of spectroscopic mass estimators in light of the second data release from Gaia is massive. The number of stars in spectroscopic surveys with accurate parallaxes has dramatically increased, and include a significant number of red giants. However to then derive ages and distances for these stars, a Bayesian isochrone pipeline needs to be applied using the spectroscopic estimate of mass. This paper expands the potential of a spectroscopic mass estimator to ages and distances  using the Bayesian Mass, Age, and Distance Estimator (MADE). The estimator is an artificial neural network (ANN), which has been demonstrated to be highly flexible in reproducing non-linear relations \citep{vangelder+14}. It is trained on stars with asteroseismology data as well as astrometric, photometric, and spectroscopic data, for which masses, ages, and distances are estimated using a Bayesian isochrone pipeline. The trained ANN completely replaces the need for stellar models. The ANN then calculates posterior predictive distributions for mass, age, and distance that account for measurement uncertainties in new data, and uncertainties in model parameters. 
All the code for the routines in the paper can be cloned from \url{https://github.com/payeldas/MADE}.

Section \ref{sec:package} introduces the components of MADE and Section \ref{sec:bnn} describes how we develop the new Bayesian spectroscopic mass, age, and distance estimator. Section \ref{sec:data} presents an application of MADE to observations, Section \ref{sec:conc} summarizes the work and looks towards future data releases.

\section[]{The MADE approach}\label{sec:package}
Here we introduce the components of MADE and the notation used to represent inputs, outputs, and model parameters.

\subsection[]{MADE components}\label{ssec:madecomp}
The core of MADE is a {\sc Python} class that builds a Bayesian ANN given training data, and then applies this ANN to unseen data. Here, the training sample comprises red giants  (i.e. $\log g < 3.5$) for which there are asteroseismology estimates for the current mass as well as parallaxes from astrometry, photometry, and spectroscopy. A Bayesian isochrone pipeline is applied to the training sample to generate posterior distributions for the desired outputs of mass, age, and distance. The ANN then learns the relationship between these outputs and input astrometric, photometric, and spectroscopic data. 
Given unseen astrometric, photometric, and spectroscopic input data of a red giant star, the ANN is used to generate predictive posterior distributions for the mass, age, and distance (see Section \ref{sec:bnn}). In this way, the ANN completely replaces the reliance on isochrones for estimating mass, age and distance from the set of inputs.

\begin{table*}
  \centering
   \caption{Random variables in MADE.\label{tab:rv}. The predicted inputs and outputs in the training sample, and the predicted inputs for new stars, also have measured equivalents that are denoted with a tilde. The measured quantities are given by the predicted quantities convolved with uncertainties.}
   \begin{tabular}{lll}
   	\hline
 	Component &Notation		&Description\\	
 	\hline
     Isochrone pipeline observed RVs &$v_j^i$ &$j^{\mathrm{th}}$ property of $i^{\mathrm{th}}$ star predicted by the isochrone pipeline\\
     &$\tilde{\rho}_j^i$         &Uncertainty in $j^{\mathrm{th}}$ measured property of $i^{\mathrm{th}}$ star \\
     Isochrone pipeline unobserved RVs &$\phi^i$ &Parameters of $i^{\mathrm{th}}$ star predicted by the isochrone pipeline\\
     &GP &Milky Way model that provides the prior on parameters of the isochrone pipeline\\
 	ANN observed RVs (training) &$x_j^i$        &$j^{\mathrm{th}}$ predicted input of $i^{\mathrm{th}}$ star in training sample\\
     &$y_j^i$         &$j^{\mathrm{th}}$ predicted output of $i^{\mathrm{th}}$ star in training sample\\
     &$u_j^i$         &$j^{\mathrm{th}}$ predicted property of $i^{\mathrm{th}}$ star in training sample\\
     &$\tilde{\sigma}_j^i$         &Uncertainty in $j^{\mathrm{th}}$ measured property of $i^{\mathrm{th}}$ star in training sample\\
     ANN observed RVs (prediction)&$x_{\mathrm{N},j}^i$        &$j^{\mathrm{th}}$ predicted input for $i^{\mathrm{th}}$ new star\\
 	  &$y_{\mathrm{N},j}^i$        &$j^{\mathrm{th}}$ predicted output for $i^{\mathrm{th}}$ new star\\
     &$\tilde{\sigma}_{\mathrm{N},j}^i$  &Uncertainty on the $j^{\mathrm{th}}$ measured input for the $i^{\mathrm{th}}$ new star\\
     ANN unobserved RVs &$\theta$ &Parameters of the ANN\\
     &$\hat{\theta}$ &Priors on parameters of the ANN\\
 	\hline
   \end{tabular}
 \end{table*}

\subsection[]{Random variables in MADE}\label{ssec:maderv}

In Bayesian statistics, everything (i.e. input labels, outputs, and model parameters) is considered to be a random variable (RV),  i.e. is described by a distribution rather than a single value. A distinction is made between an observed RV (i.e. has predicted and measured values that are linked through a likelihood depending on a measurement uncertainty), and an unobserved RV (i.e. model parameters), which is assigned a prior. The RVs in MADE are given in Table \ref{tab:rv} and explained in more detail below.

A number of observed RVs and model parameters are associated with the Bayesian isochrone pipeline used to generate the training set (Section \ref{ssec:trainingset}). The predicted observed RVs are the properties of the star $i$ generated by the isochrones and distance model
\begin{equation}
v^i = (m^i, H^i, (J-K_s)^i,\varpi^i),
\end{equation}
where $m$ is the current mass, $H$ is the $H$-band magnitude, $J-K_s$ is a colour, and $\varpi$ is the parallax. The accompanying measured observed RV stellar properties are denoted by  $\tilde{v}^i$. These are associated with measurement uncertainties, $\tilde{\rho}^i$. The parameters, $\phi^i$, of the star $i$ estimated from the Bayesian isochrone pipeline are given by
\begin{equation}
\phi^i = (\mathcal{M}^i,\tau^i,\MH^i,s^i),
\end{equation}
where $\mathcal{M}$ is the {\it initial} mass, $\tau$ is the age, $\MH$ is the metallicity, and $s$ is the distance. The priors on these parameters are informed by the Galaxy Prior (GP), which is a model for the Milky Way (Section \ref{sssec:galaxyprior}).

The observed RVs of the ANN (Section \ref{ssec:bnn}) consist of those associated with the training sample and those associated with new data. $x_j^i$ represents the $j^{\mathrm{th}}$ predicted input of the $i^{\mathrm{th}}$ repeated unit (which in practice is a star) in the training sample. $y_j^i$ represents the $j^{\mathrm{th}}$ predicted output of the $i^{\mathrm{th}}$ star in the training sample. Measurements of the predicted inputs and outputs are  $\tilde{x}_j^i$ and $\tilde{y}_j^i$ respectively. We further introduce $u_j^i$ and $\tilde{u}_j^i$, which are the predicted and measured properties of the star in the training sample respectively, where `properties' combines input and outputs into a single vector. We associate the measured properties of the star in the training sample with a measurement uncertainty $\tilde{\sigma}_j^i$. Predicted inputs and outputs for new stars are denoted by $x_{\mathrm{N},j}^i$ and $y_{\mathrm{N},j}^i$, with measurements of the input labels denoted by $\tilde{x}_{\mathrm{N},j}^i$. The measurement uncertainty on the outputs is $\tilde{\sigma}_{\mathrm{N}}^i$ and model uncertainty on the predicted outputs is $\sigma_{\mathrm{N}}^i$. We assume all observed RVs of the ANN are scaled by the mean and standard deviation of the corresponding variable of the training sample (i.e. subtract the mean and then divide by the standard deviation). The unobserved RVs are the parameters of the ANN, $\theta$, and its priors $\hat{\theta}$.

\section[]{A Bayesian spectroscopic mass, age, and distance estimator}\label{sec:bnn}
Here we discuss how we (i) build a training set for the estimator using a Bayesian isochrone pipeline, (ii) train the ANN using a Bayesian method, and (iii) use the ANN to generate posterior predictive distributions. The connection between these three components is illustrated in Figure \ref{fig:graphicalmodel}.

\subsection[]{Building a training set}\label{ssec:trainingset}

We select a training sample for which there are current mass constraints from asteroseismology in addition to parallax constraints from astrometry, photometry, and spectroscopy. We then apply a Bayesian isochrone pipeline, similar to many in the literature \citep[e.g.][]{burnett+10,carlin+15,mcmillan+17} to estimate the posterior distributions of mass, age, and distance for the stars in the training sample. The main difference however with these pipelines is that we do not estimate the line-of-sight (LOS) extinction as an additional output here. It is an output with an irregular dependence on Galactic location and therefore difficult for an ANN to capture without a sufficiently large training sample. Most of the asteroseismology data we use currently only probes one LOS. We leave this exercise of predicting the LOS extinction to a future project. We instead make a point estimate of the LOS extinction in each photometric band (see below for details).

\subsubsection{The Bayesian isochrone pipeline}\label{sssec:isochrone}

\begin{figure}
	\includegraphics[scale=0.75]{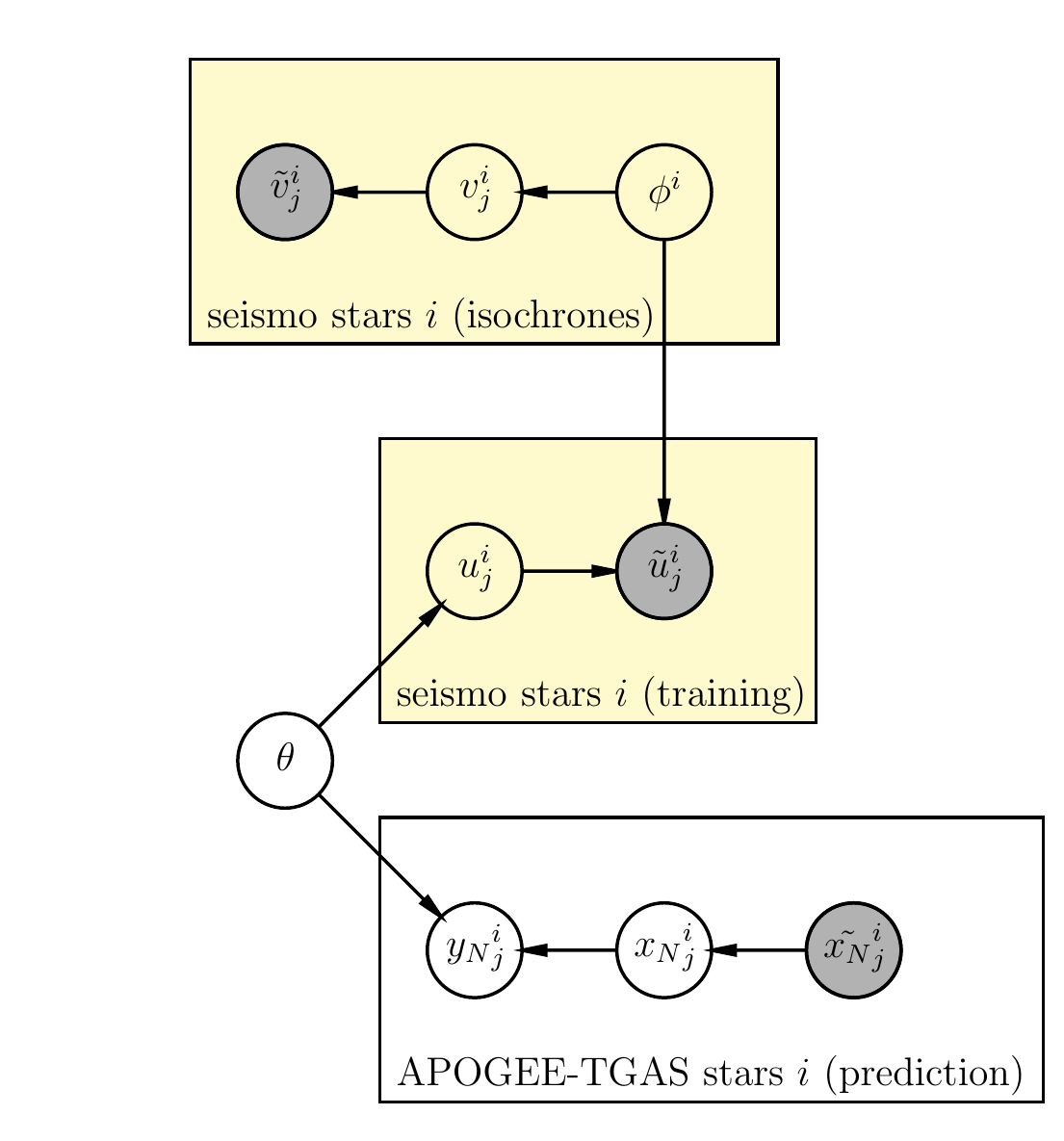}
	\caption{A graphical model of the relationships between RVs in the development of the training set using the Bayesian isochrone pipeline with the seismo stars (top panel), the training of the ANN with the seismo stars (middle panel), and the application of the ANN to the APOGEE-TGAS stars (bottom panel). Each panel represents a joint probability, where arrows indicate a correlation. Filled circles indicate observed RVs and open circles indicate unobserved RVs, i.e. parameters of the ANN or the isochrone model. Variables within a box are repeated, and therefore represent a product of probabilities. The yello highlights the training components. \label{fig:graphicalmodel}}
\end{figure}

Applying Bayes' law to each star gives
\begin{equation}\label{eqref:bayes_modpars_individstar}
p(\phi^i|\tilde{v}^i,l^i,b^i) =\frac{p(\tilde{v}^i|\phi^i,l^i,b^i) p(\phi^i|l^i,b^i)}{p(\tilde{v}^i)},
\end{equation} 
where $(l^i,b^i)$ are the predicted sky positions of star $i$ in Galactic coordinates (which we assume to be the same as the observed sky positions) and $p(v^i)$ is an unimportant normalization. The first panel in Figure \ref{fig:graphicalmodel} shows the correlations between the variables in the Bayesian isochrone pipeline and how the outputs feed into the training of the ANN (Section \ref{ssec:bnn}). The apparent magnitudes of star $i$ are extinction-corrected using the state-of-the-art \texttt{Combined15} map compiled by \cite{bovy+16b} in the \texttt{mwdust} package. This is a composite three-dimensional map of the integrated extinction. It was constructed from a number of extinction maps produced using stellar photometry but with differing sky coverage. The core is constructed from the maps of \cite{marshall+06} and \cite{green+15}. The remaining parts of the map are filled in using the map of \cite{drimmel+03}, which is normalized to the map of \cite{schlegel+98}.

The likelihood of the star's observed properties, $p(\tilde{v}^i|\phi^i,l^i,b^i)$, is assumed to be the product of the separate likelihoods. Each likelihood is represented by a Gaussian distribution
\begin{equation}
G(\tilde{v}_j^i,v_j^i,\tilde{\rho}_j^i) = \frac{1}{\sqrt{2\pi}\tilde{\rho}_j^i}\exp\left(-\frac{(\tilde{v}_j^i-v_j^i)}{2(\tilde{\rho}_j^i)^2}^2\right),
\end{equation}
Thus
\begin{equation}
p(\tilde{v}^i|\phi^i,l^i,b^i) = \prod_j G(\tilde{v}_j^i,v_j^i,\tilde{\rho}_j^i).
\end{equation}

\noindent The stellar properties, $v_j^i$, are predicted from the model parameters, $\phi^i$ using:
\begin{equation}
\begin{split}
(v_0^i,v_1^i,v_2^i) &= \mathcal{I}(\phi^i)\\ 
v_4^i &= 1/\phi_3^i\\
\end{split}
\end{equation}
where $\mathcal{I}$ denotes the isochrones and the second line refers to the distance model. When we refer to the Bayesian isochrone pipeline, we allude to both the isochrones and the distance model. Here we employ PARSEC isochrones v1.1 \citep[$\eta=0.2$, ][]{bressan+12} evaluated for 57 metallicities ranging between -2.192 and 0.696, and 353 ages ranging between $\log_{10}\tau = 6.60$ and 10.12 (i.e. a spacing of $\Delta \log_{10}\tau=0.01$) for which we create a dictionary of interpolants in {\sc Python} that returns luminosity, $\log g$, $T_{\mathrm{eff}}$ and apparent magnitudes given the metallicity, age, and mass of a star.

\begin{table}
 \centering
  \caption{Parameters of the four-component Milky Way model. \label{tab:priorpars}}
  \begin{tabular}{lll}
  	\hline
	Component		&Parameter 			   			&Value \\	
	\hline	
	Bulge			&$\mu_{\MH,1}$/dex	   			&-0.3\\
					&$\sigma_{\MH,1}$/dex   		&0.3\\
					&$\mu_{\tau,1}$/Gyr	   			&5.0\\
					&$\sigma_{\tau,1}$/Gyr  		&5.0\\		
					&$q$							&0.5\\
					&$\gamma$					    &0.0\\
					&$\delta$						&1.8\\
					&$r_0$/kpc						&0.075\\
					&$r_t$/kpc						&2.1\\
	Thin disc		&$\mu_{\MH,2}$/dex	   			&0.0\\
					&$\sigma_{\MH,2}$	   			&0.2\\		
					&$R_{\mathrm{d},2}/$kpc 		&2.6\\
					&$z_{\mathrm{d},2}/$kpc 		&0.3\\
	Thick disc		&$\mu_{\MH,3}$/dex	   			&-0.6\\
					&$\sigma_{\MH,3}$/dex   		&0.5\\
                    &$\mu_{\tau,3}$/Gyr	   			&10.\\
					&$\sigma_{\tau,3}$/Gyr   		&2.\\
					&$R_{\mathrm{d},3}/$kpc 		&3.6\\
					&$z_{\mathrm{d},3}/$kpc 		&0.9\\
	Stellar halo	&$\mu_{\MH,4}$/dex	   			&-1.6\\
					&$\sigma_{\MH,4}$/dex   		&0.5\\
					&$\mu_{\tau,4}$/Gyr	   			&11.0\\
					&$\sigma_{\tau,4}$/Gyr  		&2.0\\	
	\hline	
	\end{tabular}
\end{table}
\subsubsection{The Milky Way prior}\label{sssec:galaxyprior}
We base the prior $(\phi^i|l^i,b^i)$ on the Milky Way model of \cite{binney+14} but add a component for the bulge (the superscript $i$ is omitted in the following)
\begin{equation}
\begin{split}
p(\mathcal{M},\tau,\MH,s|l,b) = s^2\epsilon(\mathcal{M})\sum_{k=1}^{4}p_k(\MH)p_k(\tau)p_k(R,z),
\end{split}
\end{equation}
where $k=1,2,3,4$ correspond to a bulge, thin disc, thick disc, and stellar halo respectively, the $s^2$ terms accounts for the Jacobian of the transformation of spatial coordinates, and $\epsilon(\mathcal{M})$ is the initial mass function (IMF). We consider that of \cite{kroupa+93}   
\begin{equation}
	\epsilon(\mathcal{M}) = 
	\begin{cases}
		0.035\mathcal{M}^{-1.5} &\mathrm{if} \, 0.08 \leq \mathcal{M/M_{\odot}} < 0.5\\
		0.019\mathcal{M}^{-2.2} &\mathrm{if} \, 0.5 \leq \mathcal{M/M_{\odot}} < 1.0\\
		0.019\mathcal{M}^{-2.7} &\mathrm{if} \, \mathcal{M/M_{\odot}} \geq 1.0\,.
	\end{cases}
\end{equation}
The spatial prior for the bulge component is based on \cite{dehnen+06} and the metallicity prior approximately reflects the distributions found in the Abundances and Radial velocity Galactic Origins Survey (ARGOS) of the bulge \citep{ness+13}. We update the truncated age prior of \cite{binney+14} for the thick and thin discs, which assigns zero probability to young, high-latitude distant stars, to a smooth one. We also adopt a Gaussian for stellar halo ages, based on the work of \cite{jofre+11}. For the bulge we assume a broad age distribution based on \cite{bensby+13}. 

In summary, for the four components we assume,\\
\noindent Bulge ($k=1$):
\begin{equation}
\begin{split}
p_1(\MH) &= G(\MH,\mu_{\MH,1},\sigma_{\MH,1}),\\ 
p_1(\tau) &= G(\tau,\mu_{\tau,1},\sigma_{\tau,1}),\\
p_1(R,z) &\propto \frac{(1+m)^{(\gamma-\delta)}}{m^{\gamma}}\exp[-(mr_0/r_t)^2],\\
\mathrm{where}\,m(R,z) &= \sqrt{(R/r_0)^2 + (z/qr_0)^2}.
\end{split}
\end{equation}
\noindent Thin disc ($k=2$):
\begin{equation}
\begin{split}
p_2(\MH) &= G(\MH,\mu_{\MH,2},\sigma_{\MH,2}),\\ 
p_2(\tau) &\propto
\begin{cases}
	\exp(\frac{\tau}{8.4}) 
    &\mathrm{if} \, \tau/\mathrm{Gyr} \le 8 \\
    2.6\exp\left(-0.5\frac{(\tau-8)^2}{1.5^2}\right) 
    &\mathrm{if} \, \tau/\mathrm{Gyr} > 8 
\end{cases}\\
p_2(R,z) &\propto \exp\left(-\frac{R}{R_{\mathrm{d},2}}-\frac{|z|}{z_{\mathrm{d},2}}\right).
\end{split}
\end{equation}

\noindent Thick disc ($k=3$):
\begin{equation}
\begin{split}
p_3(\MH) &= G(\MH,\mu_{\MH,3},\sigma_{\MH,3}),\\
p_3(\tau) &= G(\tau,\mu_{\tau,3},\sigma_{\tau,3}),\\ 
p_3(R,z) &\propto \exp\left(-\frac{R}{R_{\mathrm{d},3}}-\frac{|z|}{z_{\mathrm{d},3}}\right).
\end{split}
\end{equation}
\noindent Stellar halo ($k=4$):
\begin{equation}
\begin{split}
p_4(\MH) &= G(\MH,\mu_{\MH,4},\sigma_{\MH,4}),\\ 
p_4(\tau) &\propto \, G(\tau,\mu_{\tau,4},\sigma_{\tau,4}) \\\
p_4(R,z) &\propto r^{-3.39}.
\end{split}
\end{equation}
The parameters of the prior are given in Table \ref{tab:priorpars}. The thin disc is normalized to have a local density of 0.04 $M_{\odot}$pc$^{-3}$ \citep{bovy17}. The thick disc and stellar halo are normalized so that the ratios of their local densities with that of the thin disc are 0.04 and 0.005, respectively \citep{bland+16,bovy17}. Finally, the bulge disc is normalized to have a central density of 35.45 $M_{\odot}$pc$^{-3}$ \citep{robin+12}.  An overall prior is imposed that constrains metallicities and ages to the range covered by the isochrones.

We calculate $p(\phi^i|u^i)$ directly on a grid of all 353 ages, $\tau^i$,  all metallicities, $\MH^i$, lying within $3\sigma$ of the measured metallicity, 2000 initial masses, $\mathcal{M}^i$, ranging between the minimum and maximum mass of the relevant isochrone, and 30 distances, $s^i$, based on a linear grid of parallaxes ranging between 3$\sigma$ below and above the measured parallax. $\tau^i$ and $\MH^i$ define the isochrone and $\mathcal{M}^i$ defines the position along the isochrone. The position on the isochrone returns a prediction of mass, $m^i$, colours, and absolute magnitudes, i.e. the predicted stellar properties $v^i$. The choice of $s^i$ tells us the apparent magnitudes. We calculate the first and second moments of the logarithm of age, $\log_{10}\tau$, metallicity, [M/H], logarithm of the current mass, $\log_{10}m$, and the distance modulus, $\mu$ as our outputs and output uncertainties, e.g.
\begin{equation}
\begin{split}
\langle \log_{10}\tau \rangle = \Big(\int \,\mathrm{d}\phi\,\log_{10}\tau \,p(u|\phi,l,b) p(\phi|l,b)\Big) \Big/\\ \Big(\int \mathrm{d}\phi\,p(u|\phi,l,b)p(\phi|l,b)\Big).
\label{eq:moments}
\end{split}
\end{equation}

\def\layersep{3.5cm}
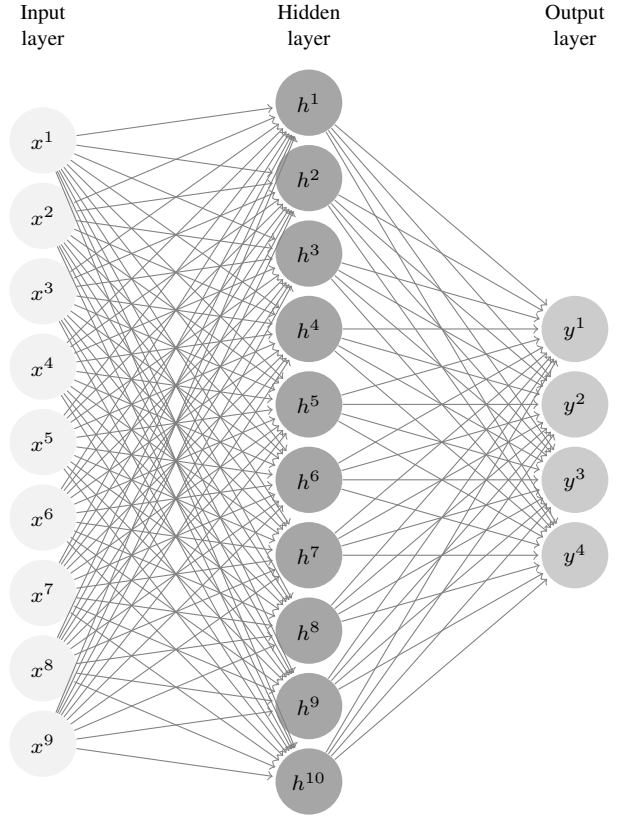
\begin{figure}
	\begin{tikzpicture}[shorten >=1pt,->,draw=black!50, node distance=\layersep]
    	\tikzstyle{every pin edge}=[<-,shorten <=1pt]
    	\tikzstyle{neuron}=[circle,fill=black!25,minimum size=25pt,inner sep=0pt]
	    \tikzstyle{input neuron}=[neuron, fill=gray!10];
    	\tikzstyle{output neuron}=[neuron, fill=gray!40];
    	\tikzstyle{hidden neuron}=[neuron, fill=gray!70];
    	\tikzstyle{annot} = [text width=4em, text centered]
		\foreach \name / \y in {1,...,9}
        \node[input neuron] (I-\name) at (0,-\y) {$x^{\name}$};

    	\foreach \name / \y in {1,...,10}
        \path[yshift=0.5cm]
        	    node[hidden neuron] (H-\name) at (\layersep,-\y cm) {$h^{\name}$};

    	\node[output neuron,right of=H-4] (O) {$y^1$};
    	\node[output neuron,right of=H-5] (1) {$y^2$};
        \node[output neuron,right of=H-6] (2) {$y^3$};
    	\node[output neuron,right of=H-7] (3) {$y^4$};
        
    	\foreach \source in {1,...,9}
        	\foreach \dest in {1,...,10}
            	\path (I-\source) edge (H-\dest);

    	\foreach \source in {1,...,10}
        	\path (H-\source) edge (O);
		\foreach \source in {1,...,10}
        	\path (H-\source) edge (1);
        \foreach \source in {1,...,10}
        	\path (H-\source) edge (2);
		\foreach \source in {1,...,10}
        	\path (H-\source) edge (3);
            
    	\node[annot,above of=H-1, node distance=1.cm] (hl) {Hidden layer};
    	\node[annot,left of=hl] {Input layer};
    	\node[annot,right of=hl] {Output layer};
	\end{tikzpicture}
	\caption{Architecture of ANN assuming $n_{\mathrm{in}}=9$ (light gray), $n_{\mathrm{out}}=4$ (middle gray), and one hidden layer with $n_{\mathrm{hid}}=10$ neurons (dark gray).\label{fig:ann}}
\end{figure}
\subsection[]{The Bayesian ANN}\label{ssec:bnn}
An ANN consists of interconnected layers of neurons, which represent linear or non-linear transformations by an `activation' function. The first layer is the input layer comprising the same number of neurons as the number of inputs, $n_{\mathrm{in}}$. The central layers are hidden layers, each with potentially a different number of neurons per hidden layer, $n_{\mathrm{hid}}$. The final layer is the output layer with the same number of neurons as the number of outputs, $n_{\mathrm{out}}$.

\subsubsection[]{ANN architecture}\label{sssec:ann_arch}
The universality theorem\footnote{A visualization of this theorem can be found at http://neuralnetworksanddeeplearning.com/chap4.html} tells us that even ANNs with a single hidden layer can be used to approximate any continuous function to any desired precision (given a sufficient number of hidden neurons). We expect a continuous mapping from one finite space to another and therefore consider a simple ANN architecture that contains a single hidden layer. 

In a feed forward ANN, only neurons in adjacent layers are connected to one another. We assume linear activation functions for the input and output layers and a $\tanh$ activation function for the hidden layer, which maps variables ranging from $-\infty$ to $\infty$ to a domain extending between -1 to 1. One can imagine easily replicating non-linear relations by stacking shifted and differentially stretched versions of this sigmoidal function. Therefore the predicted outputs $y_i$ are calculated from the predicted inputs $x_i$ by
\begin{equation}
y^i = (w_{\mathrm{h},\mathrm{out}}\tanh(\mathbf{w}_{\mathrm{h},\mathrm{in}}x^i+b_{\mathrm{h},\mathrm{in}}))+b_{\mathrm{h},\mathrm{out}},
\end{equation}
where $\mathbf{w}_{h,\mathrm{in}}$ is a $n_{\mathrm{hid}}\times n_{\mathrm{in}}$ matrix of weights, $b_{h,\mathrm{in}}$ is a length-$n_{\mathrm{hid}}$ vector of biases, $w_{h,\mathrm{out}}$ is a length-$n_{\mathrm{hid}}$ vector of weights, and $b_{h,\mathrm{out}}$ is a length-$n_{\mathrm{out}}$ vector of biases. This architecture has $n_{\theta} = n_{\mathrm{hid}}(n_{\mathrm{in}}+2) + n_{\mathrm{out}}$ model parameters. Figure \ref{fig:ann} shows the structure of the ANN in the case of $n_{\mathrm{hid}}=10$.

\subsubsection[]{Training the ANN}\label{sssec:anntraining}

The posterior distributions of model parameters, $\theta$, can be estimated using Bayes' law
\begin{equation}\label{eq:bayes_modpars}
  p(\theta|\mathbf{\tilde{u}}) =\frac{p(\mathbf{\tilde{u}}|\theta) p(\theta)}{p(\mathbf{\tilde{u}})},
\end{equation} 
where $p(\mathbf{\tilde{u}}|\theta)$ is the joint likelihood of the measured stellar properties given the model parameters, $p(\theta)$ is the prior on the model parameters, and $p(\mathbf{\tilde{u}})$ is the distribution of the measured stellar properties. $p(\mathbf{\tilde{u}})$ is the same for every model and can be ignored. The correlations between the variables are illustrated in the middle panel of Figure \ref{fig:graphicalmodel}. The Figure also shows how the Bayesian isochrone pipeline feeds into the training of the ANN, and how the parameter distributions determined at this stage are used to generate posterior predictive distributions in Section \ref{sssec:annpp}. We do not explicitly include measurement and model uncertainties in the representation of our probabilities throughout the paper as they only appear in likelihoods. The likelihood of the star's measured properties $p(\mathbf{\tilde{u}}|\theta)$ is assumed to be the product of the likelihoods of each measured stellar property. Assuming Gaussian measurement uncertainties, we can represent each likelihood as
\begin{equation}
G(\tilde{u}_j^i,u_j^i,\tilde{\sigma}_j^i) = \frac{1}{\sqrt{2\pi}\tilde{\sigma}_j^i}\exp\left(-\frac{(\tilde{u}_j^i-u_j^i)}{2{{(\tilde{\sigma}}_j^i)^2}}\right).
\end{equation}
Thus
\begin{equation}\label{eq:lh}
p(\mathbf{\tilde{u}}|\theta) = \prod_i\prod_j G(\tilde{u}_j^i,u_j^i,\tilde{\sigma}_j^i).
\end{equation}
The predicted outputs are those generated by the ANN. The predicted inputs are initially unknown and therefore assigned Gaussian priors with zero mean and a large standard deviation of 10. We assume Gaussian priors with mean zero and standard deviation 1 for the ANN parameters. 

The No-U-Turn Sampler \citep[NUTS, i.e. adaptive Hamiltonian Monte Carlo,][]{hoffman+14} in PyMC3 is used to train the ANN on 80\% stars randomly selected from a sample of stars for which both inputs and outputs have been measured. The remaining 20\% is used as an independent test of the ANN. 

\subsection[]{Generating posterior predictive distributions}\label{sssec:annpp}

Once we have obtained posterior distributions for the parameters of the ANN, $p(\theta|\mathbf{\tilde{u}})$, we can calculate posterior predictive distributions for selected predicted stellar properties of new stars, $\mathbf{y}_N$, given the training sample and the new measured inputs, i.e. 
\begin{equation}\label{eq:pps}
p(\mathbf{y}_{\mathrm{N}}|\mathbf{\tilde{u}},\mathbf{\tilde{x}_{\mathrm{N}}}) = \int\int p(\mathbf{y}_{\mathrm{N}}|\theta)p(\theta|\mathbf{\tilde{u}})p(\mathbf{y}_{\mathrm{N}}|\mathbf{x}_{\mathrm{N}})p(\mathbf{x}_{\mathrm{N}}|\tilde{\mathbf{x}}_{\mathrm{N}})\diff\theta\diff \mathbf{x}_{\mathrm{N}},
\end{equation}
where $p(\mathbf{y}_{\mathrm{N}}|\theta)$ is the probability of the new predicted outputs given some set of model parameters, $p(\theta|\mathbf{\tilde{u}})$ is the posterior distributions of the ANN model parameters evaluated in Section \ref{sssec:anntraining}, $p(\mathbf{y}_{\mathrm{N}}|\mathbf{x_{\mathrm{N}}})$ gives the distribution of new predicted outputs given new predicted inputs, which are simply the predictions of the trained ANN. $p(\mathbf{x}_{\mathrm{N}}|\tilde{\mathbf{x}}_{\mathrm{N}})$ is the distribution of new predicted inputs given the new measured inputs, which is the likelihood as given in Equation \ref{eq:lh}. Marginalizing the product of these probabilities over the parameters of the ANN and the new inputs give the posterior predictive distributions on the new outputs. Generating the posterior predictive distributions therefore does not require one to engage with the isochrones. The correlations between the variables are illustrated in the bottom panel of Figure \ref{fig:graphicalmodel}. The Figure also shows how the parameter distributions found in the training process are linked to the new predicted outputs. 

\section[]{Application}\label{sec:data}

Here, we introduce the APOGEE-TGAS sample of stars and a further subsample for which current masses have been determined from asteroseismology. 
\begin{figure}
	\centering
	\includegraphics[scale=0.55]{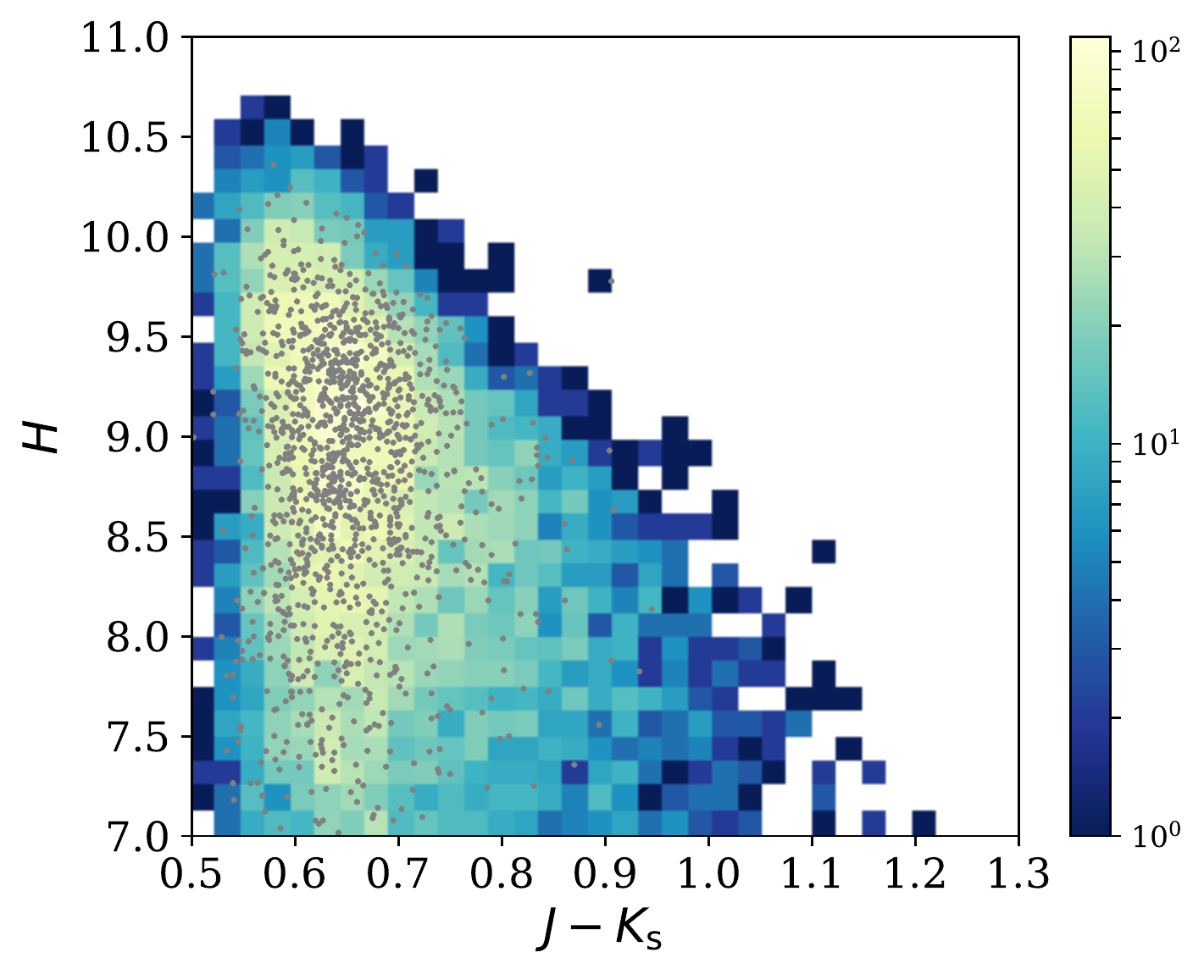}
	\caption{Colour-magnitude distribution of stars in the APOGEE-TGAS sample. The grey dots show the location of the seismo sample.\label{fig:data_colmag}}
\end{figure}

\subsection[]{The APOGEE-TGAS catalogue}\label{ssec:apogee_tgas}
APOGEE spectra are taken in the $H$-band with a resolution $R \sim 22\,500$. The first phase of the APOGEE Survey (APOGEE-1) was carried out between September 2011 and July 2014 \citep{majewski+17}. Observations were performed with the APOGEE-North spectrograph on the Sloan Foundation 2.5m Telescope of Apache Point Observatory (APO). The second phase, APOGEE-2, started in July 2014 and will be completed in the summer of 2020 \citep[][]{zasowski+17}. Observations will also be taken with the APOGEE-South spectrograph on the Ir\'{e}n\'{e}e du Pont 2.5m Telescope of Las Campanas Observatory (LCO). Data Release 14 \cite[DR14,][]{abol+18} contains $\sim 263\,000$ of mostly red giants, but with a significant contribution from red dwarf stars. APOGEE DR14 releases spectra \citep{nidever+15} and derived spectroscopic properties that include 20 individual chemical abundances \citep{garcia+16}. Surface gravities are calibrated using independent asteroseismology determinations with the Kepler mission \citep{haas+10}. Kepler is described in more detail below. 

The Gaia DR1 TGAS catalogue \citep{mich+15} provides positions, parallaxes, and proper motions for the 2.5 million Tycho-2 stars \citep{hog+00}. APOGEE DR14 includes a cross-match with these stars that results in $\sim 46\,000$ stars. We select unique stars by keeping those with the highest signal-to-noise ratio, and remove stars with unreliable abundances and those that are cluster or calibration targets. We then select stars within the designed APOGEE colour-magnitude box, i.e. $0.5\le (J-K_{\mathrm{s}})_0\le1.3$ (we imposed the upper bound) and $7.0<H<13.8$. These operations reduce the sample to 10\,074 stars. The distribution of the final sample in the space of colours and magnitudes is shown in Figure \ref{fig:data_colmag}. Assuming $\log g\le 3.5$ for giant stars \citep{hekker+11}, 10\,016 are very likely to be giants indicating a very small contamination of dwarf stars. We therefore assume the catalogue to comprise only giants.

\subsection[]{The seismo-APOGEE-TGAS (seismo) sample}\label{ssec:seismo}

The Kepler spacecraft was launched in March 2009 \citep{haas+10} and spent a little over four years monitoring the brightness of more than $150\,000$ stars in the Cygnus-Lyra region, with the primary science objective of detecting transit-driven exoplanet stars. This was succeeded by the K2 mission \citep{howell+14} in June 2014. We construct a catalogue of asteroseismology masses by combining the catalogue of \cite{vrard+16} with the first APOKASC catalogue \citep{pinson+14}. Both initially determine mode spacings in the Kepler photometric data and then apply scaling relations to determine masses and radii. 1235 of these stars also have TGAS parallaxes, creating the seismo-APOGEE-TGAS sample, or seismo sample for short. The location of these stars in the colour-magnitude diagram is also shown in Figure \ref{fig:data_colmag}. They have a similar distribution to the APOGEE-TGAS stars in this space. 

\subsection{A new Bayesian spectroscopic mass, age, and distance estimator}\label{ssec:bsmade}
\begin{figure}
	\centering
	\includegraphics[scale=0.6]{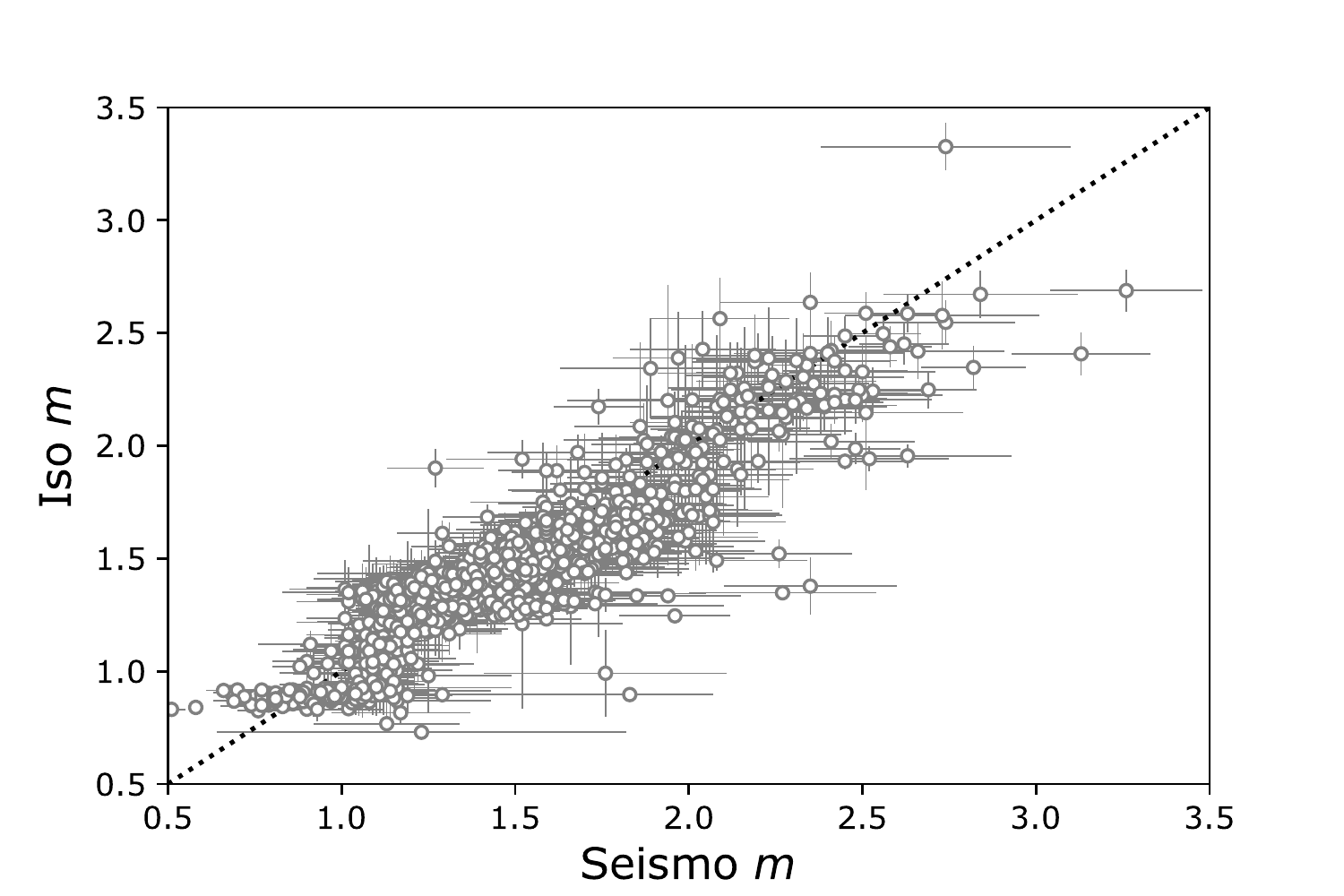}
	\caption{Masses determined from asteroseismology (seismo $m$) against masses determined from the Bayesian isochrone pipeline (iso $m$). The uncertainty in the isochrone mass is estimated from $10^{\mu_{\log_{10}m}} \sigma_{\log_{10}}m$. \label{fig:seismo_iso_mass_comp}}
\end{figure}

\begin{table}
 \centering
  \caption{Correlation between each potential combination of input and output for the ANN. The correlation is measured with the Spearman Rank Correlation Coefficient. If the associated $p$-value is less than 0.05, then we take the correlation to be significant, and the cell is coloured grey.}\label{tab:srcc}
  \begin{tabularx}{.7\columnwidth}{lXXXX}
  	\hline
	 		 &$\log_{10}\tau$    &[M/H] &$\log_{10}m$ &$\mu$\\	
	\hline	
	$H$ &\cellcolor{Gray} &\cellcolor{Gray} &\cellcolor{Gray} &\cellcolor{Gray}\\
	$J-K$ & &\cellcolor{Gray} &   &\cellcolor{Gray}\\
	$\varpi$ &\cellcolor{Gray} &\cellcolor{Gray}&\cellcolor{Gray}&\cellcolor{Gray}\\
	$\log g$ &\cellcolor{Gray} &\cellcolor{Gray}&\cellcolor{Gray}&\cellcolor{Gray}\\
	$T_{\mathrm{eff}}$&\cellcolor{Gray} &\cellcolor{Gray} &\cellcolor{Gray} &\\
    $[\mathrm{M/H}]$ &\cellcolor{Gray} &\cellcolor{Gray} &\cellcolor{Gray} &\cellcolor{Gray}\\
	$[\alpha/\mathrm{M}]$&\cellcolor{Gray} &\cellcolor{Gray} &\cellcolor{Gray}&\cellcolor{Gray}\\
    $[\mathrm{C/M}]$&\cellcolor{Gray} &\cellcolor{Gray} &\cellcolor{Gray} &\\
	$[\mathrm{N/M}]$&\cellcolor{Gray} &\cellcolor{Gray} &\cellcolor{Gray} &\\
	\hline	
\end{tabularx}
\end{table}
We apply the Bayesian isochrone pipeline described in Section \ref{sssec:isochrone} to the seismo sample of stars and determine the first and second moments of $\log_{10} \tau$, [M/H], $\log_{10} m$, and $\mu$. Figure \ref{fig:seismo_iso_mass_comp} compares the asteroseismology masses with the isochrone masses. We do expect some differences between the masses as the isochrone masses are informed by astrometric, photometric, and spectroscopic data, as well as the asteroseismology data. In general though they agree well. The uncertainty in isochrone masses is generally smaller due to a larger number of observational constraints being used, and priors at the minimum and maximum ages. The prior on age in the Bayesian isochrone pipeline results in a cut off at the lowest and maximum masses, due to the mass-age relation for red giant stars. The scaling relations used to derive the asteroseismology masses do not impose such a prior. The isochrone masses therefore systematically deviate from the asteroseismology masses for the lowest and highest masses.

As metallicity is an additional output to age, distance, and mass, we consider the following ANN outputs
\begin{equation}
y^i = (\log_{10}\tau^i,\mathrm{[M/H]}^i,\log_{10} m^i,\mu^i).
\end{equation}
To build the Bayesian spectroscopic estimator, we first investigate which inputs can be used as predictors of these outputs. Calculating the Spearman Rank Correlation Coefficient and associated $p$-value between a set of potential predictors and the desired outputs (Table \ref{tab:srcc}), we find all of them to be significantly correlated with at least one of the outputs. We therefore take the ANN inputs to be
\begin{multline}
x^i = (H^i, (J-K)^i, \varpi^i, \log g^i, \\ T_{\mathrm{eff}}^i,\mathrm{[M/H]}^i,[\alpha\mathrm{/M]}^i,\mathrm{[C/M]}^i,\mathrm{[N/M]}^i).
\end{multline}
As discussed earlier, age and mass are closely related for red giant stars, and therefore they have significant correlations with the same set of inputs. The apparent magnitude combined with parallax tells us the absolute magnitude of the star, which reflects the age, metallicity, and mass of the star. Although the distance of the star is primarily driven by its parallax, it is also modified by the absolute magnitude of the star, which is related directly to the apparent magnitude. The mass of the star also sets $\log g$, the radius of the star, and therefore $T_{\mathrm{eff}}$ of the star. The age of a star correlates with [M/H] and [C/N] as a result of both stellar evolution and Galactic chemical evolution.

Considering successful outputs from the Bayesian isochrone pipeline, and valid measurements for $x^i$, our seismo sample reduces to 1214 stars. We explore three ANN architectures each with a single hidden layer, and either 5, 10, or 15 neurons in the hidden layer. The top part of the graphical model in Figure \ref{fig:graphicalmodel} summarizes the training of the ANN.
\begin{figure*}
	\centering
	\includegraphics[scale=0.7]{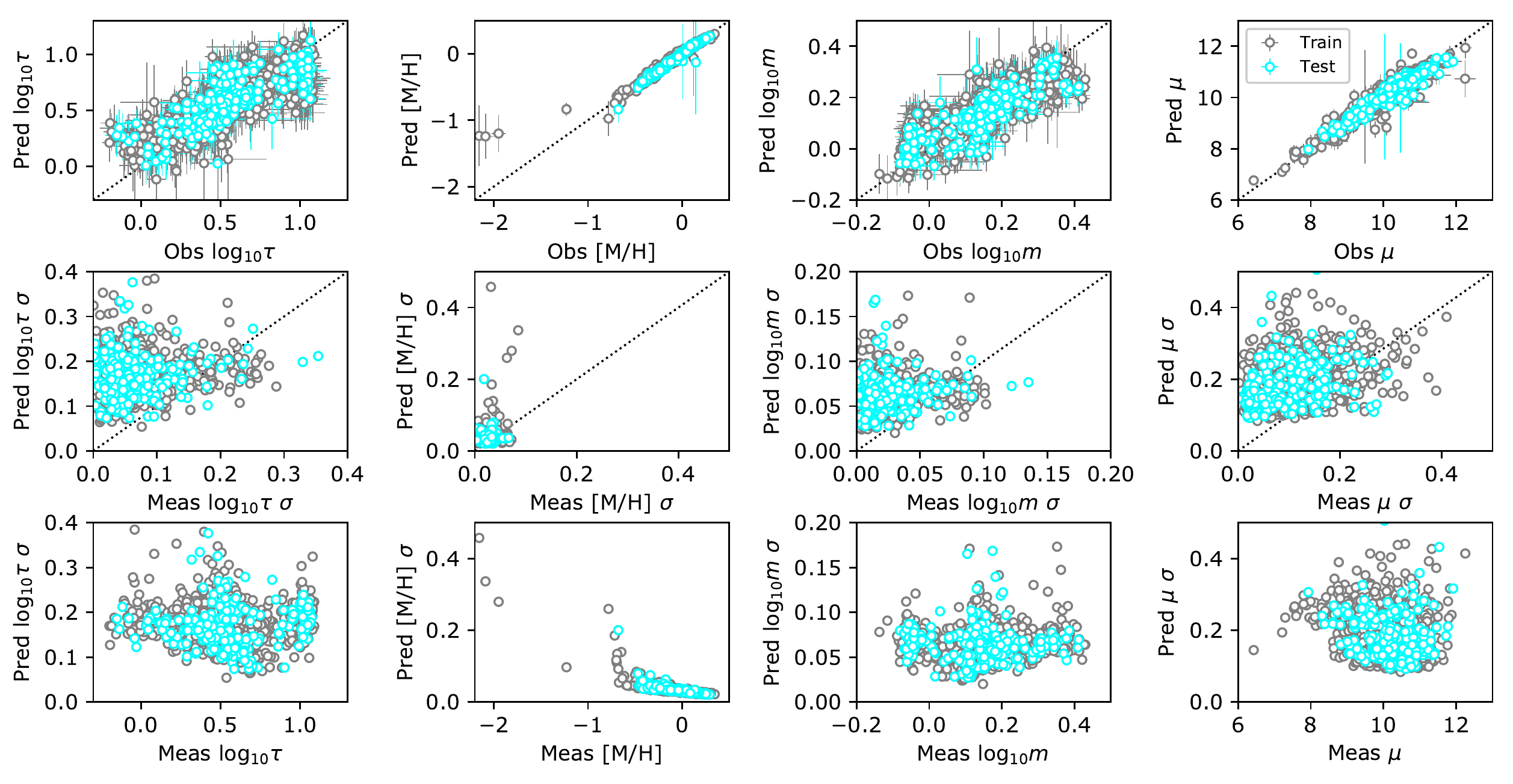}
	\caption{A comparison between predicted and measured means (top panel), predicted and measured standard deviations (middle panel), and predicted standard deviation and measured mean (bottom panel). The predicted output distributions are generated by model B and the measured output distributions are generated by the Bayesian isochrone pipeline for the training (grey) and testing (cyan) samples. \label{fig:obs_model_comp}}
\end{figure*}
\begin{figure}
	\centering
	\includegraphics[scale=0.58]{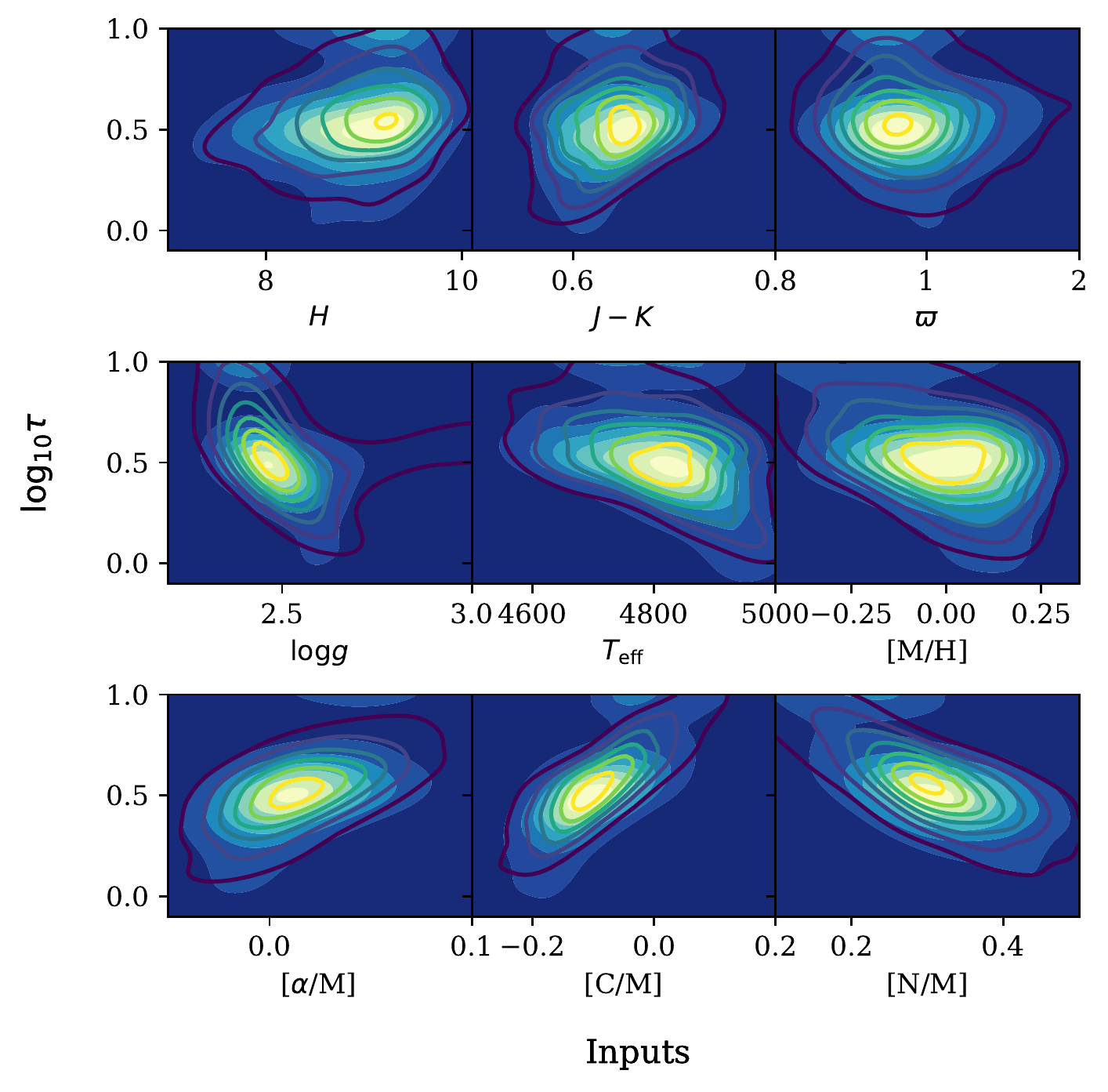}
	\caption{Contour plots of the joint distributions of a selection of the inputs (parallax, photometry, and spectroscopy) and one of the  outputs (age) for the combined training and testing sets. The measurements are shown by filled contours and the coloured contour lines present joint distributions of model B.\label{fig:obs_model_comp_2d}}
\end{figure}

\begin{table}
 \centering
  \caption{Performance of different ANN architectures: Columns (1) architecture, (2) total number of parameters in ANN (i.e. $n_{\mathrm{hid}}(n_{\mathrm{in}}+2)+n_{\mathrm{out}}$ ), (3) reduced $\chi^2$ for each output of the testing set (1-$\log_{10}\tau$, 2-[M/H], 3-$\log_{10} m$, and 4-$\mu$. The selected model is highlighted in grey.}\label{tab:chi2}
  \begin{tabular}{llllll}
  	\hline
	Architecture  &$n_{\theta}$ &$\chi^2_{\mathrm{r},1}$ &$\chi^2_{\mathrm{r},2}$ &$\chi^2_{\mathrm{r},3}$ &$\chi^2_{\mathrm{r},4}$\\	
	\hline	
	A. $n_{\mathrm{hid}}  = 5$   &59  &0.706 &0.213 &0.682 &0.793\\
    \rowcolor{Gray}
	B. $n_{\mathrm{hid}}  = 10$   &114  &0.705 &0.184 &0.719 &0.745\\
	C. $n_{\mathrm{hid}}  = 15$   &169  &0.781 &0.197 &0.785 &0.703\\	
	\hline	
\end{tabular}
\end{table}

The best ANN architecture is estimated by calculating a reduced $\chi$-squared, $\chi^2_{\mathrm{r},j}$, for each predicted output. This statistic is formulated to consider both uncertainty in the prediction from the ANN and the uncertainty in the measurements of the outputs
\begin{equation}
\chi^2_{\mathrm{r},j} = \frac{1}{n_{\mathrm{test}}}\sum_k^{n_{\mathrm{test}}}\frac{(y_{\mathrm{N},j}^{k} - \tilde{y}_{\mathrm{N},j}^{k})^2}{(\sigma_{\mathrm{N},j}^{k})^2+(\tilde{\sigma}_{\mathrm{N},j}^{k})^2},
\end{equation}
where $n_{\mathrm{test}}$ is the number of stars in the testing sample. Table \ref{tab:chi2} shows ${\chi^i_{\mathrm{r}}}^2$ for each output and for the range of explored architectures. We select model B as the best model though note that $\chi^2_{\mathrm{r}}$ is always below 1, and so all models perform well.

\begin{figure*}
	\centering
	\includegraphics[scale=0.7]{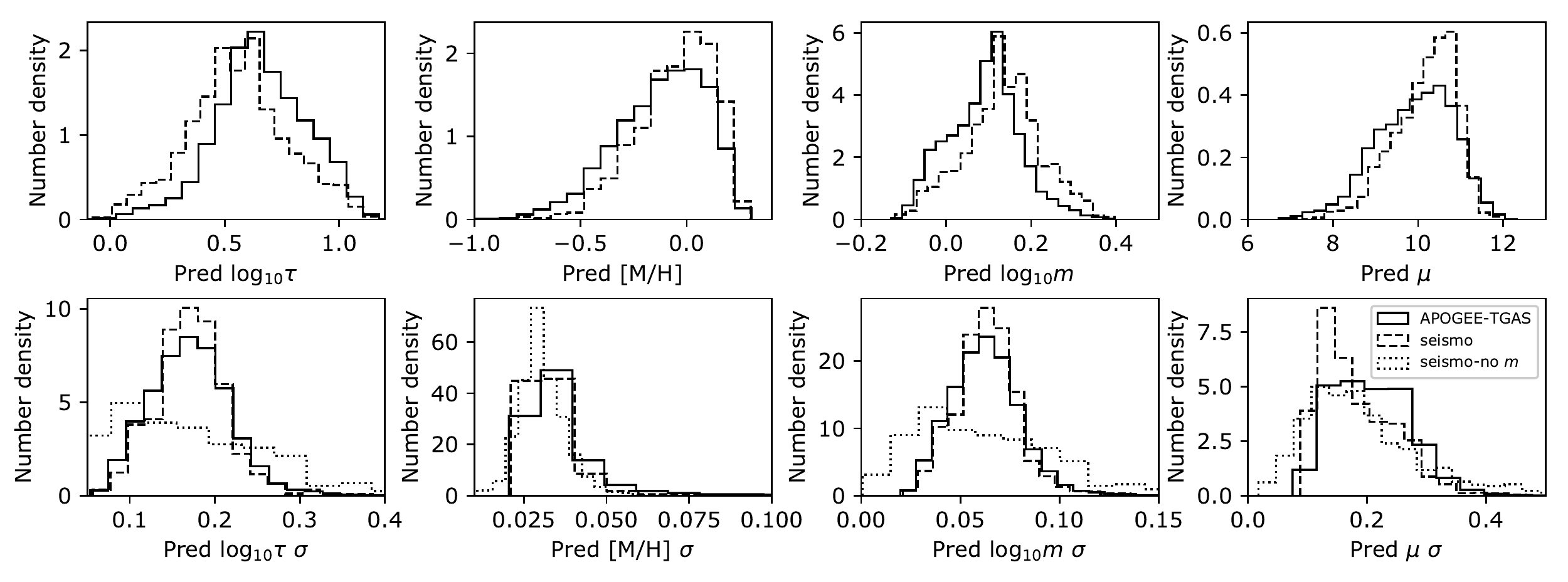}
	\caption{Histograms of the predicted mean and uncertainty in age, metallicity, mass, and distance derived by the ANN for the APOGEE-TGAS sample (solid, black), compared with that evaluated by the ANN for the seismo sample with the ANN (dashed, black). The bottom panel also shows uncertainty distributions when applying the Bayesian isochrone pipeline to the seismo sample but not using the asteroseismology mass information (dotted, black). \label{fig:apogeetgas_ann_outputs}}
\end{figure*}

Figure \ref{fig:obs_model_comp} compares the means and standard deviations of the output distributions predicted by our favoured model B against those of the measured output distributions for the combined training and testing samples. As expected, the ANN performs almost perfectly in the prediction of [M/H] and $\mu$. [M/H] is also an input to the ANN and only marginally corrected by the other input variables. $\mu$ is primarily driven by $\varpi$, with spectro-photometric variables contributing more for distant stars. For the three most metal-poor stars however, the ANN systematically overestimates the metallicity, possibly due to a lack of stars at the metal-poor end in the training sample. Age and mass are generally recovered well, but there is significantly more scatter. The lowest ages are systematically overestimated and the highest ages systematically underestimated, as a result of the effect of the age prior on generating isochrone ages and therefore masses.

The range of uncertainties recovered by the ANN is smaller than the uncertainties in the isochrone estimates. For stars with the smallest uncertainties in the isochrone estimates, the ANN estimates are generally higher. This probably reflects a lack of sufficient flexibility in the ANN architecture. However many stars with the highest isochrone uncertainties have lower ANN uncertainties, showing the powerful possibility of using the ANN to `denoise' our data.

The uncertainties on the predicted outputs are flat with the means of the observed outputs, except in the case of [M/H], for which the prediction capability gently degrades towards lower metallicities. Beyond a metallicity of -1, the ANN does not perform well.

The predicted joint probability distributions between a selection of the inputs (parallax, photometry, and spectroscopy) and one of the outputs (age) for the combined training and testing sets for model B are shown compared to the measured joint distributions in Figure \ref{fig:obs_model_comp_2d}. The plot shows that the Bayesian ANN in general does a good job of reproducing the joint distributions, except for the oldest stars where the age prior in the Bayesian isochrone pipeline forces a pile-up towards the oldest and youngest ages. 

As a final illustration of the power of the Bayesian ANN, we redetermine masses, ages, distances, and metallicities for the training and testing samples, but assume there is no asteroseismology mass information. The uncertainty distributions in the bottom panel of Figure \ref{fig:apogeetgas_ann_outputs} show that the ANN is not able to produce age and mass estimates with as low uncertainties as the Bayesian isochrone pipeline for some stars. As stated above, this can probably be improved by allowing more flexibility in the ANN. It could also be due to the isochrones giving precise mass estimates that are biased with respect to the masses that would be predicted by carbon and nitrogen only. As the ANN considers both estimates, the associated uncertainty will be larger. The ANN also performs much better than the Bayesian isochrone pipeline for many stars, partly because of the carbon and nitrogen information, and partly due to the capability of the ANN to `denoise' the noisiest isochrone outputs.

\subsection[]{New masses, ages, distances, and metallicities for APOGEE-TGAS stars}\label{ssec:distage}

Selecting the APOGEE-TGAS stars within the same input domain spanned by the seismo stars, and further cutting to limit the sample to metallicities above -1.0 reduces the sample from 10074 to 9403 stars. Applying the ANN takes just over a minute on a single core to calculate posterior predictive distributions for mass, age, distance, and metallicity. This is comparable to the time taken for the Baysian isochrone pipeline to run on a handful of stars. Figure \ref{fig:apogeetgas_ann_outputs} shows the distributions in the mean and standard deviations predicted for the APOGEE-TGAS stars. The age distribution peaks at 4-5 Gyr (and mass at around 1.3 $M_{\odot}$), the metallicity peaks at solar metallicity, and the distance modulus at 10.5 mag (a distance of 1.3 kpc). The uncertainty distributions show that mass can be estimated to almost always better than 10\%, and ages to between 10 and 25\%. The uncertainties on metallicity and distance modulus are, as expected, small.

\begin{figure*}
	\centering
	\includegraphics[scale=0.55]{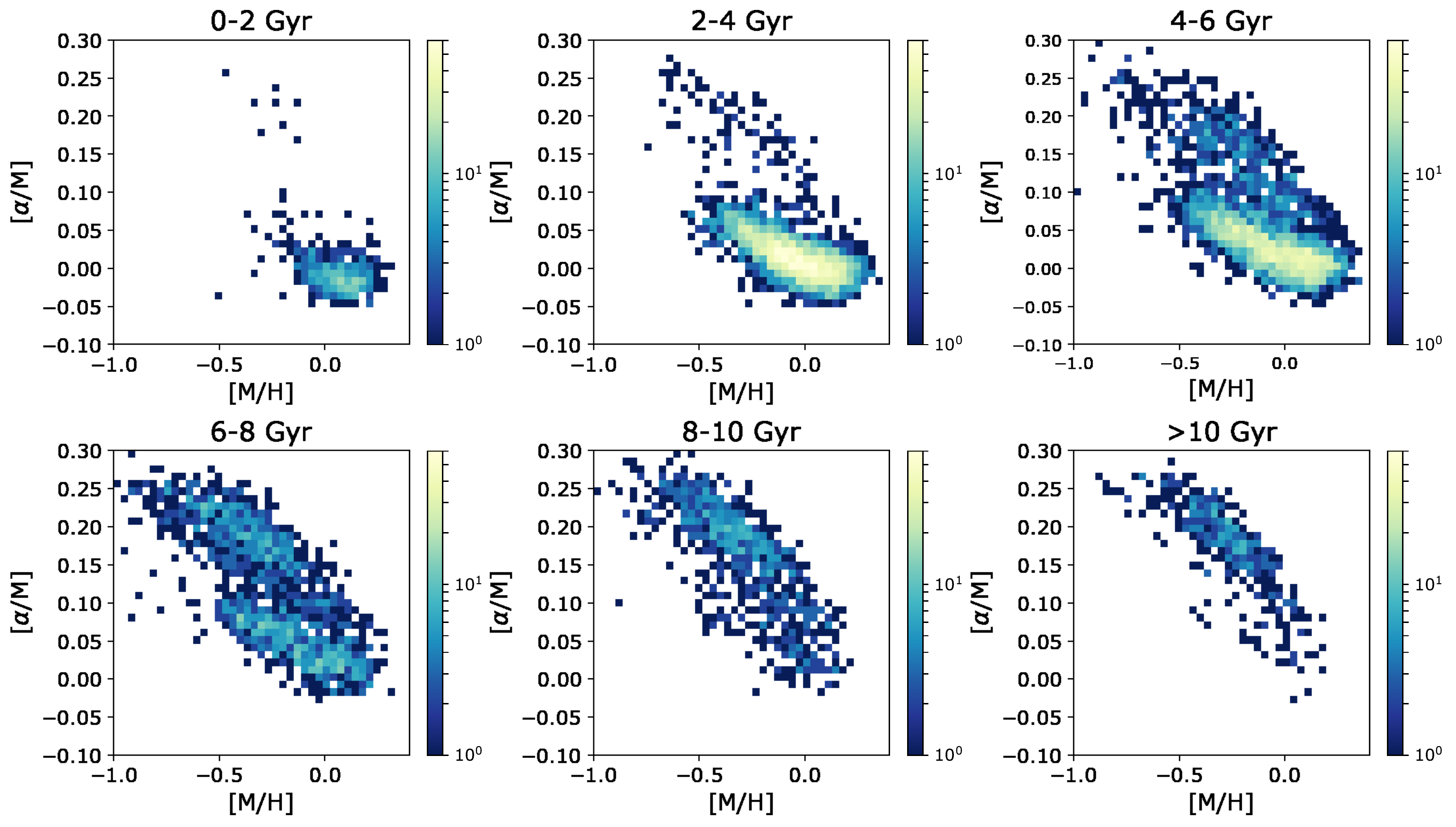}
	\caption{Distribution of APOGEE-TGAS stars in the $\MH$-$\aM$ plane for different age slices.\label{fig:mh_am_age_dist}}
\end{figure*}
To check whether the ages correlate with abundances as expected, we consider the density of stars in age slices in the $\MH$-$\aM$ plane (Figure \ref{fig:mh_am_age_dist}). Between 0-2 Gyr, most stars are relatively metal rich with a low $\alpha$ abundance. From 2-4 Gyr, the sequence of metal-rich, low-$\alpha$ stars has become more evident, and a sequence of high-$\alpha$, metal-poor stars appears. The metal-rich, low-$\alpha$ component starts to fade for larger ages and has almost disappeared for stars older than 10 Gyr. The metal-poor, high-$\alpha$ component is most dominant between 6-8 Gyr. Two sequences have been discussed in the literature several times \citep[e.g.][]{hayden+14}.  The high-$\alpha$, metal-poor stars can be considered to be the `thick disc', formed $\sim$8 Gyr ago \citep{haywood+13}, while the low-$\alpha$ metal-rich stars can be considered to be the `thin disc', that has an extended star formation history. The thin disc is therefore expected to have stars ranging from those born in the present time to those as old as $\sim$8 Gyr, while the thick disc stars are all expected to be $\sim$8 Gyr old. This is very similar to what we see. It should be noted that uncertainties in age can cause the stars to scatter between the age slices. This could result in thick-disc stars appearing at younger ages, and thin-disc stars appearing at older ages.

\section[]{Conclusions and future outlook}\label{sec:conc} 
We demonstrate the potential for ANNs for precise predictions of masses, ages, and distances for red giants using the Bayesian spectroscopic Mass, Age, and Distance Estimator. A Bayesian isochrone pipeline was applied to a training sample of giant stars with apparent magnitudes and colours, $\log g$ and $T_{\mathrm{eff}}$ from APOGEE, asteroseismology masses from Kepler, and parallaxes from TGAS to derive posterior distributions for mass, age, distance, and metallicity. We then use this sample, complementing the input data with APOGEE carbon and nitrogen abundance estimates, to train a Bayesian ANN to learn the relationship between the inputs and our desired outputs. The ANN on average reproduces the isochrone estimates for mass, age, distance, and metallicity with similar uncertainties. For training stars with high output uncertainties, the ANN reduces the uncertainty probably as a result of using the extra information in carbon and nitrogen.

The ANN is then used to estimate posterior predictive distributions for masses, ages, distances, and  metallicities for the whole APOGEE-TGAS sample. This procedure takes about a minute, and so is significantly quicker than applying the Bayesian isochrone pipeline. The fractional uncertainties on mass are at worst $\sim 10\%$ and on age range between 10 to 25\%.

The regression presented here is inspired by similar regressions developed by \cite{martig+16} and \cite{ness+16} to estimate spectroscopic masses and ages for the same stars in APOGEE. \cite{martig+16} developed a polynomial regression between carbon abundance, nitrogen abundance, $\log g$, $T_{\mathrm{eff}}$, $\FeH$, and mass using 1475 giants with asteroseismology mass estimates from Kepler. They predicted stellar masses with fractional uncertainties of about 14\% and ages with fractional uncertainties of about 40\%. \cite{ness+16} use the same asteroseismology mass estimates with the {\it Cannon}, a sophisticated data-driven model that uses the full spectrum. This also estimated ages with fractional uncertainties of about 40\%. The ANN developed here uses spectral parameters rather than the full spectrum. However, it employs a Bayesian approach unlike the methods of \cite{martig+16} and the Cannon, deriving posterior distributions for the model parameters and ANN outputs. It is able to estimate age, distance, and metallicity as well as mass, therefore omitting the expensive Bayesian isochrone pipeline. 

The work presented here is exploratory and there is significant room for improvement. In light of parallaxes that have just arrived for over a billion stars from Gaia DR2 and the second APOKASC catalogue, a much larger training sample can be built. New isochrones \citep[][PARAM]{PARAM} exist that provide $\Delta\nu$ and $\nu_{\mathrm{max}}$, removing the reliance on scaling relations that may have been one of the sources of discrepancies between the asteroseismology masses and the isochrone masses. The age prior in the Bayesian isochrone pipeline may need to be examined in more detail. The ANN architecture may lack sufficient flexibility as it cannot reproduce the small uncertainties achieved for some stars with the Bayesian isochrone pipeline. Finally, the potential for the ANN to also estimate the LOS extinction is a powerful future extension.

To summarize, the method presented here can calculate posterior predictive distributions for masses, ages, distances, and metallicities quickly for a large number of stars. Several spectroscopic surveys now have parallaxes for all stars in their samples. These include APOGEE \citep[DR14][]{abol+18}, the Radial Velocity Experiment survey \citep[RAVE DR5,][]{kunder+17}, the Galactic Archaeology with HERMES survey \citep[GALAH DR2,][]{buder+18}, Gaia-ESO, and The Large sky Area Multi-Object Fiber Spectroscopic Telescope survey \citep[LAMOST DR4,][]{he+16}. These surveys and many more future surveys will greatly benefit from the application of MADE to efficiently calculate masses, ages, distances, and metallicities.

\section*{Acknowledgements}
The research leading to these results has received funding from the ERC under the European Union's Seventh Framework Programme (FP7/2007-2013)/ERC grant agreement no. 321067 and the `Extracting science from Galaxy surveys' project on the Science and Technology Facilities Council (STFC) consolidated grant `Astrophysics at Oxford' (grant agreement no. ST/N000919/1). 

This project was inspired by the 2016 NYC Gaia Sprint, hosted by the Center for Computational Astrophysics at the Simons Foundation in New York City. PD would also like to thank Professor Suzanne Aigrain at the University of Oxford for her expertise regarding the construction of graphical models and Bayesian statistics.

We thanks GitHub for providing free private repositories for educational use. We are also indebted to a number of \textsc{Python} packages such as Astropy \citep{astropy} and packages in the \texttt{scipy} ecosystem \citep{scipy,matplotlib,ipython,pandas}. It is a pleasure to thank the Galactic Dynamics group in Oxford for many fruitful discussions, in particular Professor James Binney for a thorough read of the manuscript.

This work has made use of data from the European Space Agency (ESA) mission
{\it Gaia} (\url{https://www.cosmos.esa.int/web/gaia/iow_20180316}), processed by the {\it Gaia}
Data Processing and Analysis Consortium (DPAC,
\url{https://www.cosmos.esa.int/web/gaia/dpac/consortium}).

This publication makes use of data products from the Two Micron All Sky Survey, which is a joint project of the University of Massachusetts and the Infrared Processing and Analysis Center/California Institute of Technology, funded by the National Aeronautics and Space Administration and the National Science Foundation.

Funding for the Sloan Digital Sky Survey IV has been provided by the Alfred P. Sloan Foundation, the U.S. Department of Energy Office of Science, and the Participating Institutions. SDSS-IV acknowledges support and resources from the Center for High-Performance Computing at the University of Utah. The SDSS web site is www.sdss.org.

\bibliographystyle{mnras}
\bibliography{biblio}

\appendix

\label{lastpage}

\end{document}